\documentclass[%
 aip,
 jmp,%
 amsmath,amssymb,
 reprint,%
]{revtex4-1}

\usepackage{graphicx}
\usepackage{dcolumn}
\usepackage{bm}
\usepackage{xcolor}

\begin{document}

\preprint{AIP/123-QED}

\title[Backflashes \textcolor{black}{from} fast-gated avalanche photodiodes in quantum key distribution]{Backflashes \textcolor{black}{from} fast-gated avalanche photodiodes in quantum key distribution}

\author{A. Koehler-Sidki}
 \affiliation{ Toshiba Research Europe Ltd, Cambridge Research Laboratory,
208 Cambridge Science Park, Milton Road, Cambridge, CB4 0GZ, United Kingdom}
 \affiliation{ Engineering Department, University of Cambridge,
9 J. J. Thomson Avenue, Cambridge CB3 0FA, United Kingdom}
\author{J. F. Dynes}%
 \email{james.dynes@crl.toshiba.co.uk}
\affiliation{
 Toshiba Research Europe Ltd, Cambridge Research Laboratory, 208 Cambridge Science Park, Milton Road, Cambridge, CB4 0GZ, United Kingdom
}%

\author{T. K. Para\"{i}so}
\affiliation{
 Toshiba Research Europe Ltd, Cambridge Research Laboratory, 208 Cambridge Science Park, Milton Road, Cambridge, CB4 0GZ, United Kingdom
}%

\author{M. Lucamarini}
\affiliation{
 Toshiba Research Europe Ltd, Cambridge Research Laboratory, 208 Cambridge Science Park, Milton Road, Cambridge, CB4 0GZ, United Kingdom
}%
\author{A. W. Sharpe}
\affiliation{
 Toshiba Research Europe Ltd, Cambridge Research Laboratory, 208 Cambridge Science Park, Milton Road, Cambridge, CB4 0GZ, United Kingdom
}%
\author{Z. L. Yuan}
\affiliation{
 Toshiba Research Europe Ltd, Cambridge Research Laboratory, 208 Cambridge Science Park, Milton Road, Cambridge, CB4 0GZ, United Kingdom
}%
\author{A. J. Shields}
\affiliation{
 Toshiba Research Europe Ltd, Cambridge Research Laboratory, 208 Cambridge Science Park, Milton Road, Cambridge, CB4 0GZ, United Kingdom
}%

\date{\today}

\begin{abstract}
InGaAs single-photon avalanche photodiodes (APDs) are key enablers for high-bit rate quantum key distribution. However, the deviation of such detectors from ideal models can open side-channels for an eavesdropper, Eve, to exploit. The phenomenon of backflashes, whereby APDs reemit photons after detecting a photon, gives Eve the opportunity to passively learn the information carried by the detected photon without the need to actively interact with the legitimate receiver, Bob. Whilst this has been observed in slow-gated detectors, it has not been investigated in fast-gated APDs where it has been posited that this effect would be lessened. Here, we perform the first experiment to characterise the security threat that backflashes provide in a GHz-gated self-differencing APD using the metric of information leakage. We find that, indeed, the information leakage is lower than that reported for slower-gated detectors and we show that its effect on the secure key rate is negligible. We also relate the rate of backflash events to the APD dark current, thereby suggesting their origin is the InP multiplication region in the APD.
%
\end{abstract}

\pacs{Valid PACS appear here}
\maketitle

Quantum key distribution (QKD) promises information theoretic security that is guaranteed by the laws of physics \cite{bennett1984}. This property has spurred significant efforts in this research area, culminating in a number of field trials \cite{Peev09,Sasaki11,Dynes12,Mao:18,Dixon17,Bunandar18,Sun18}. With the recent deployment of QKD outside of the lab, avalanche photodiodes (APDs) have presented themselves as the most promising single-photon detectors due to their ability to operate at room temperature \cite{Comandar14}, excellent detection efficiency \cite{Comandar16} and short dead-times \cite{Yuan18}.

Whilst perfectly secure in theory, deviations of components from their ideal behaviour can create security loopholes. Detectors are the most vulnerable devices in a QKD system as they are exposed through the optical channel and therefore are the most accessible component to Eve. One example exists in the form of the faked-state attack \cite{Makarov05}, of which the most notable implementation is the blinding attack. Demonstrations of this attack have been presented on a variety of individual detectors and systems \cite{Lydersen_hacking_10, Gerhardt11}, although several of these have only been possible due to inappropriate operation rather than a genuine security weakness \cite{Yuan10, Yuan11}.

The aforementioned attacks are all examples of Eve actively interacting with the QKD system, both by measuring Alice's qubits and then illuminating Bob's detectors. This presents a significant chance of her presence being detected. It has been shown that APDs are susceptible to \textcolor{black}{emitting light after a detection, known as backflashes} \cite{Meda17, Pinheiro18, Marini17, Shi17, Kurtsiefer01}. Backflashes can then allow Eve to act in a more passive way and thus ascertain which of Bob's detectors has clicked without having to interact with any components in the QKD system. However, no studies have yet been performed on fast-gated detectors that are used in state-of-the-art QKD systems \cite{Yuan18}. Whilst it has been suggested that faster gating, resulting in shorter gates and subsequently avalanches with less charge, would result in fewer backflashes \cite{Meda17}, this hypothesis has not been experimentally verified.


In this paper, we present the first study on backflashes in GHz-gated self-differencing APDs, key enablers in high bit rate QKD \cite{Yuan18}. Our finding support the hypothesis that faster gating, resulting in narrower gates and smaller avalanche charges, results in fewer backflashes. Using the technique in Ref. \onlinecite{Meda17}, we quantify the information leakage and find it to be 0.5\%, which is an order of magnitude lower than the value measured for a MHz-gated detector. Such a low information leakage has a negligible effect on the secure key rate.

To determine the potential vulnerability of a fast-gated APD, we perform a simple experiment. An InGaAs/InP APD is chosen as the device under test. It is thermoelectrically cooled to --30$^{\circ}$C where the breakdown voltage is 62.16V. When driven with a constant DC bias of 59.66V and a peak-to-peak 1 GHz AC signal of 10V \textcolor{black}{with 50\% duty cycle corresponding to 500 ps `ON' and `OFF' times, respectively}, the APD exhibits a detection efficiency of 17\% at a wavelength of 1550~nm, a dark count probability of $1.9 \times 10^{-6}$ and an afterpulse probability of 5\%.

\begin{figure}[htbp!]
    \centering
    \includegraphics[width=0.48\textwidth]{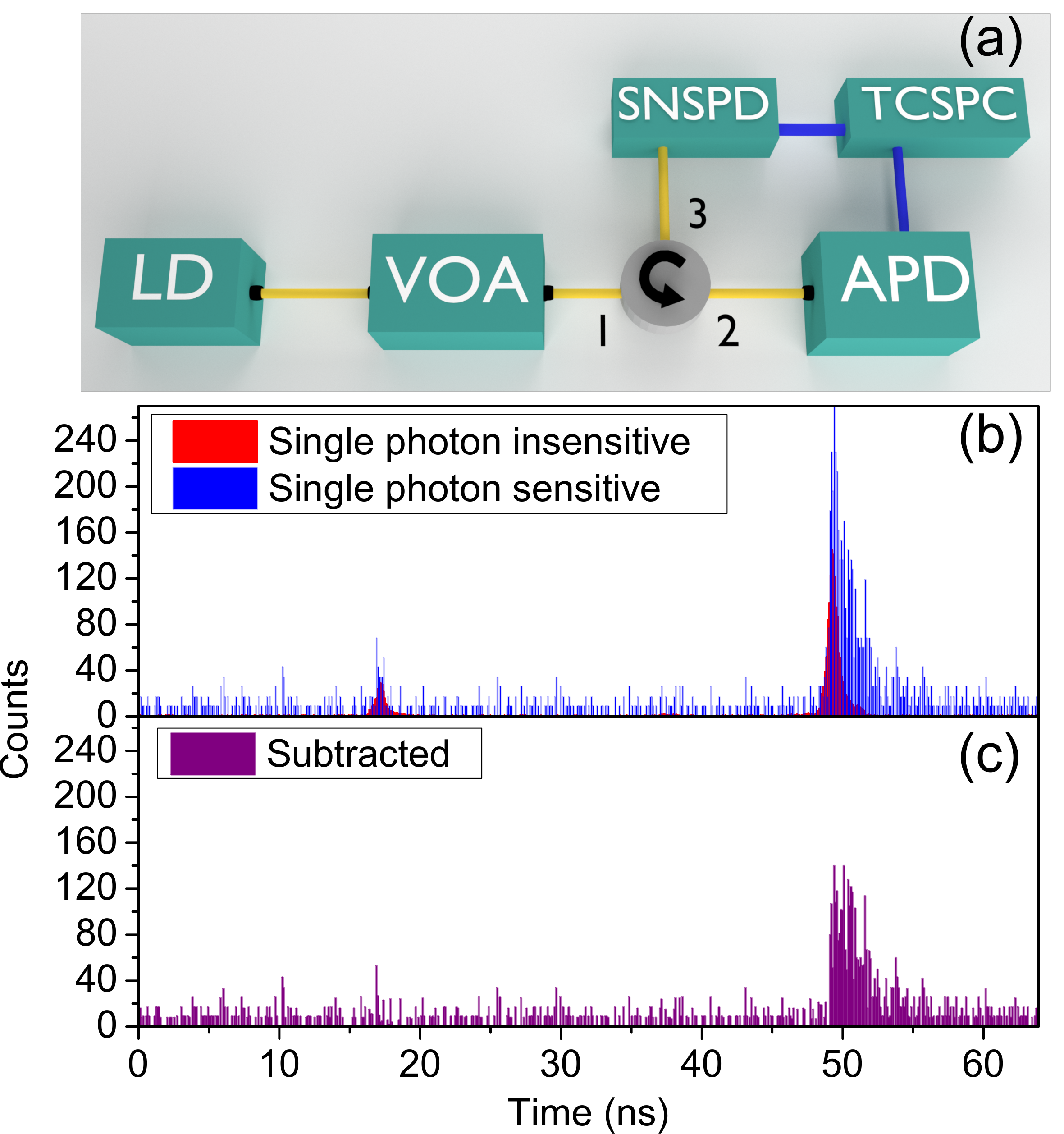}
    \caption{(a) Schematic of the experiment used to investigate APD backflashes. LD: laser diode; VOA: variable optical attenuator; SNSPD: superconducting single photon detector; TCSPC: time-correlated single-photon counter. (b) Histograms of the detection events on the SNSPD when the APD is illuminated with a 0.1 photons/pulse \textcolor{black}{where the total measurement time is 10 seconds}. \textcolor{black}{The x-axis refers to the effective delay with respect to the laser trigger pulse.} The APD is biased under two different DC biases: single photon sensitive (blue bars) and single photon insensitive (red bars). (c) Subtracted histogram with backreflections removed, leaving only backflashes}
    \label{fig:setup_hists}
\end{figure}

For investigating the effect of backflashes on the security of QKD, the APD is illuminated with a 1550~nm pulsed laser diode with a \textcolor{black}{pulse width of approximately 30 ps and} repetition frequency of $1/64$ of the APD gating frequency (15.625 MHz). \textcolor{black}{We use this laser repetition frequency to only illuminate every 64th gate as this allows us to mitigate the addition of afterpulses when determining the number of legitimate APD counts. If a faster frequency were used, afterpulses could raise the APD detections and thus artificially lower the information leakage.} The flux is controlled using a variable optical attenuator. We illuminate the APD with 0.1~photons/pulse, a flux typical for QKD, at the start of the APD gate. \textcolor{black}{The reasons for the placement of the pulse in this temporal location are twofold. Firstly, this simulates the behaviour of the legitimate users, as the detection efficiency is greatest at the start of the gate. Secondly, placing the pulse at the start of the gate gives the avalanches the longest time to grow and therefore provides a maximum value of the backflash probability and is therefore the more conservative estimate of information leakage.} The light enters port 1 of a circulator and port 2 is connected to the APD. Emitted backflashes then re-enter the circular and exit via port 3, after which they are measured with a superconducting nanowire detector (SNSPD). The detected APD counts and backflashes are interpreted with a time-tagging single-photon counter. This is illustrated in Fig.~\ref{fig:setup_hists}(a).

In an ideal case, any light detected by the SNSPDs can be attributed to backflashes. However, in the optical path are also detected and can artificially raise the SNSPD count rate. An example of this is shown in the histogram of SNSPD detection events with the APD DC and AC disabled, see red bars in Fig.~\ref{fig:setup_hists}(b). The peak features at approximately 17 and \textcolor{black}{49 ns} can be attributed to backreflections \textcolor{black}{from the APD surface and connector between APD and circulator} and they dominate the SNSPD detection events when the APD is single-photon insensitive, as shown as the red bars in the same figure. The blue bars corresponding to backflashes are reasonably uniformly distributed across the histogram, with the exception of the second backreflected peak. At this point of approximately 49 ns, the blue bars have a much larger amplitude (around 100 rather than 40) which suggests that this peak corresponds to reflection from the APD surface itself and that the backflashes are strongly correlated with APD detection events.

To quantify the effect of backflashes on QKD security, we use the metric of information leakage, defined in Ref. \onlinecite{Meda17} as

\begin{align}
    P_{L} = \frac{N_{B}}{N_{A}\eta_{det}\eta_{ch}},
    \label{eq:PL}
\end{align}

\noindent where $N_{B}$ is the number of detected backflashes (neglecting backreflections and dark counts), $N_{A}$ is the number of detected valid APD counts (i.e neglecting dark counts), $\eta_{det}$ is the detection efficiency of the monitoring detector (80\% for the SNSPD used), and $\eta_{ch}$ is the channel loss between the APD under test and the monitoring detector, measured to be 0.78.

\begin{figure}[htbp!]
    \centering
    \includegraphics[width=0.48\textwidth]{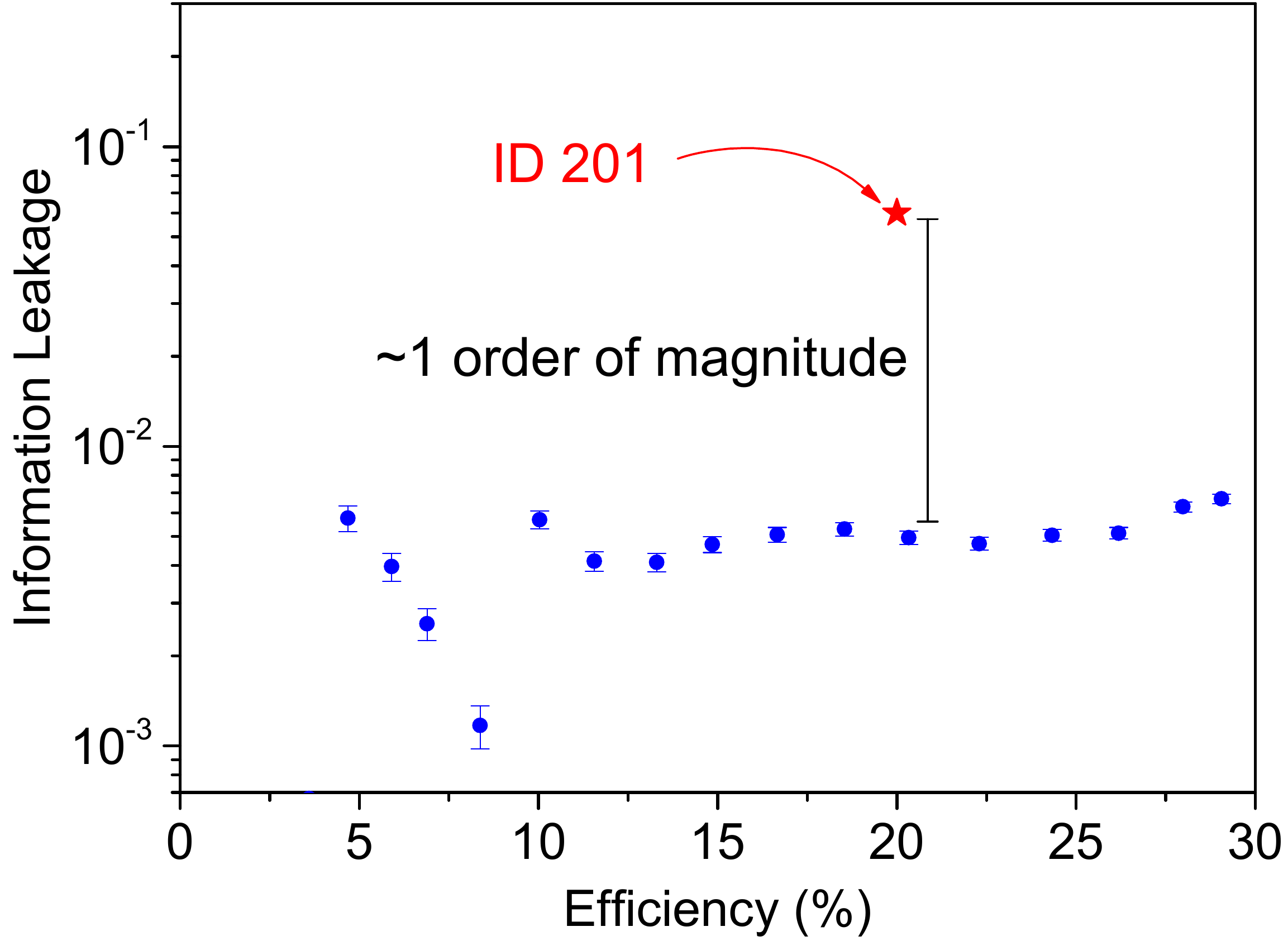}
    \caption{\textcolor{black}{Information leakage plotted as a function of the APD single-photon detection efficiency. The red star indicates the corresponding information leakage for a commercially available APD, ID 201, reported in the literature \cite{Meda17}. The detector under test exhibits an order of magnitude smaller information leakage, supporting the hypothesis that faster-gated APDs emit fewer backflashes.}}
    \label{fig:P_L}
\end{figure}

In order to obtain a true measure of the information leakage, it was necessary to isolate the backreflections. A simple technique for this is simply to neglect them in post processing. This was done by subtracting the SNSPD histogram with the APD turned off, so that only backflashes were measured, shown in Fig.~\ref{fig:setup_hists}(c). This large peak also at around 49 ns supports the hypothesis given above that the backflashes are correlated with APD clicks. \textcolor{black}{Measurements were performed for different detection efficiencies by varying the DC bias to the APD, and the subsequent information leakage then calculated and plotted as a function of detection efficiency in Fig.~\ref{fig:P_L} alongside the value measured in Ref. 17 for the ID 201 detector. The APD detection efficiency was determined at each point using the technique outlined in \cite{Comandar15}.} 

The data appears initially very noisy at low efficiency. This is due to the SNSPD count rate being similar to its dark count rate, which suggests the rate of backflashes is very low. \textcolor{black}{We note that it was not possible to extended the measurement time to smooth out the statistics due to the instability of the APD's temperature over time.} The data then appears much smoother from an efficiency of 10\% as the rate of backflashes increases. As the information leakage remains more or less constant from then on, this suggests the relationship between backflashes and APD counts is linear. By comparing this to the \textcolor{black}{ID 201} detector, we see an order of magnitude improvement in the information leakage, which supports the hypothesis that shorter gates will emit fewer backflashes.

Using the value for information leakage, which is a direct measurement of Eve's information, we can derive a new secure key rate in the presence of backflashes. This has been partially investigated in Ref. \onlinecite{Pinheiro18} where the authors approach the derivation of the key rate from a photon number splitting perspective and treat the information leakage as `tagged' bits, but originating from Bob rather than Alice \cite{Gottesman04,Inamori07}. However, the authors in Ref. \onlinecite{Pinheiro18} assume the backflash probability, and therefore information leakage, remains constant over all distances, which means they obtain a very pessimistic estimate for the secure key rate. \textcolor{black}{This is because they use the conditional backflash probability (i.e. the probability of a backflash if there is an APD click), whereas the raw, absolute backflash probability would have been more appropriate.} In reality, as the information leakage is dependent on an APD click, the APD click probability should also be incorporated into this analysis so that the key rate is affected by the same proportion, regardless of distance. We use a modified version of the key rate given in Ref. \onlinecite{Pinheiro18} considering single-photon BB84 as follows

\begin{equation}\label{eq:skr_bk}
  R \geq q P_{click} \big [ (1-P_{L})  \left \{1-h(e)\right \} - \left \{fh(e) \right \} \big ]
\end{equation}

\noindent where $q$ is the basis choice probability, $P_{click}$ is the probability of a click on a detector, $P_{L}$ is the information leakage (defined in equation~\ref{eq:PL}), $h(x)$ is the binary Shannon entropy, \textcolor{black}{$e$ is the quantum bit error rate} and $f$ is the error correction efficiency. It is interesting to note that by simply multiplying the information leakage term by the click probability in the key rate definition from Ref. \onlinecite{Pinheiro18}, thereby including a dependence of the backflash probability on the APD detection probability, that equation reduces to equation~\ref{eq:skr_bk}.

\begin{figure}[htbp!]
    \centering
    \includegraphics[width=0.48\textwidth]{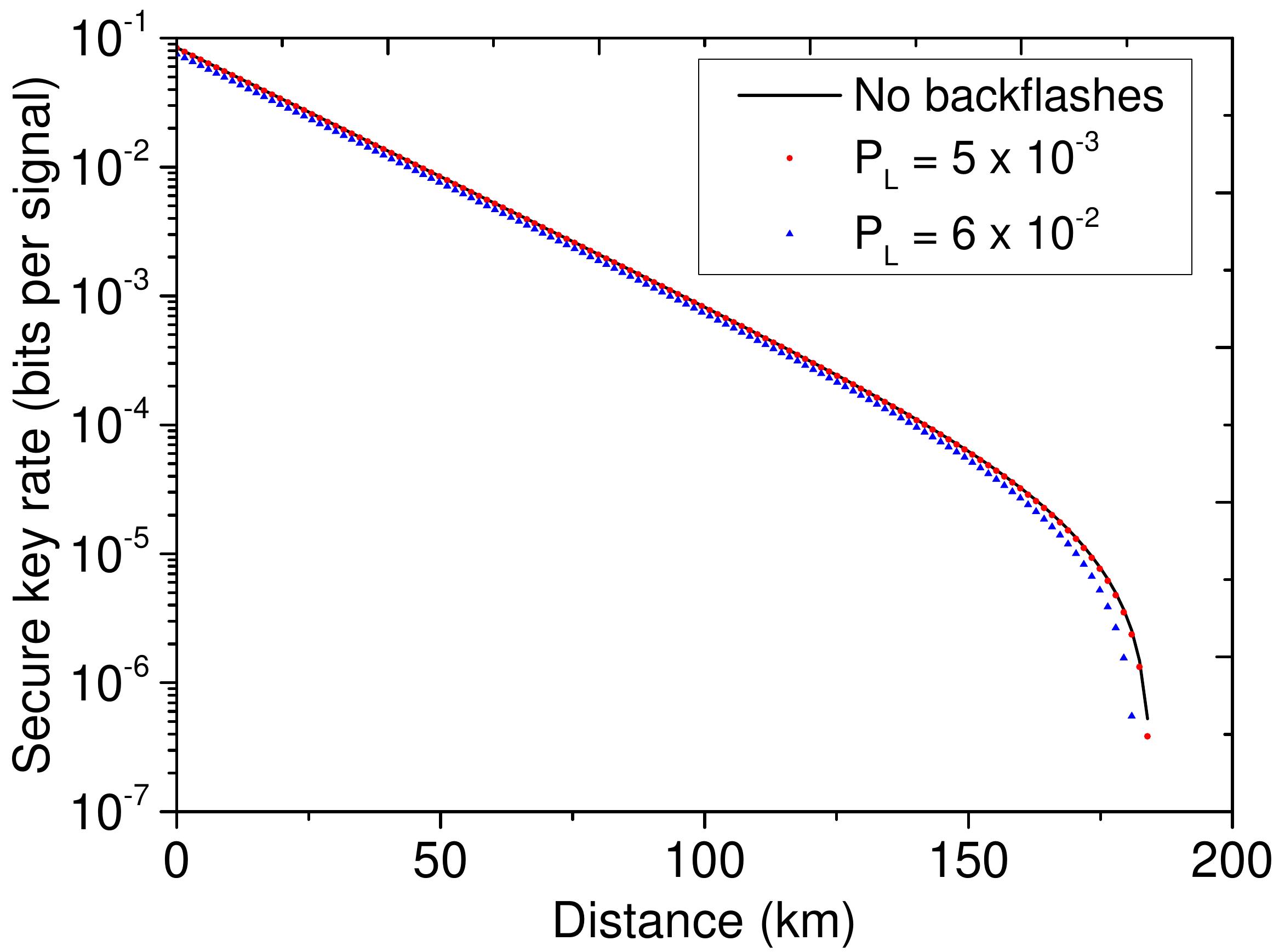}
    \caption{Secure key rate plotted in the absence of backflashes, with the measured information leakage and previous state-of-the-art. Even with $P_{L}=6\%$, the effect on the key rate is negligible, as the term $P_{L}$ gives the exact amount by which the key rate is reduced.}
    \label{fig:SKRinfo_leakage}
\end{figure}

Using detector characteristics from this study we plot the key rate as a function of distance for several values of information leakage, namely zero, $5 \times 10^{-2}$, which was the previous state-of-the-art and $5 \times 10^{-3}$, as measured in our own setup, as shown in Fig.~\ref{fig:SKRinfo_leakage}.

As an information leakage of 0.5\% has a negligible effect on the key rate, an isolator would not be needed as a countermeasure since even with a very low insertion loss of 0.2 dB, it would have a greater impact on the key rate. This result provides strong evidence that backflashes are not a significant threat to QKD, even for slower gated detectors where the information leakage is potentially larger. \textcolor{black}{We note, however, that characterising the spectrum of the backflashes is also important for enforcing this point in order to more accurately determine the information leakage. Whilst this has been partially explored in previous studies \cite{Shi17, Marini17, Pinheiro18}, these have not corrected for the spectral response of the measurement apparatus. We believe this is an important avenue for future work, not only from a security perspective but also to shed light on the precise origin of backflashes within APDs.}

\textcolor{black}{Whilst we have shown that backflashes have a small effect on the secure key rate, they can still pose a security risk. As shown in Ref. \onlinecite{Meda17}, the temporal profile of backflashes appears to be unique for different APDs. This can provide Eve with information on the detectors used by Bob, allowing her to use a tailored attack that is dependent on the type of APD in Bob's system. Therefore, the use of an isolate may still be desirable as a countermeasure.}

\begin{figure}[htbp!]
    \centering
    \includegraphics[width=0.48\textwidth]{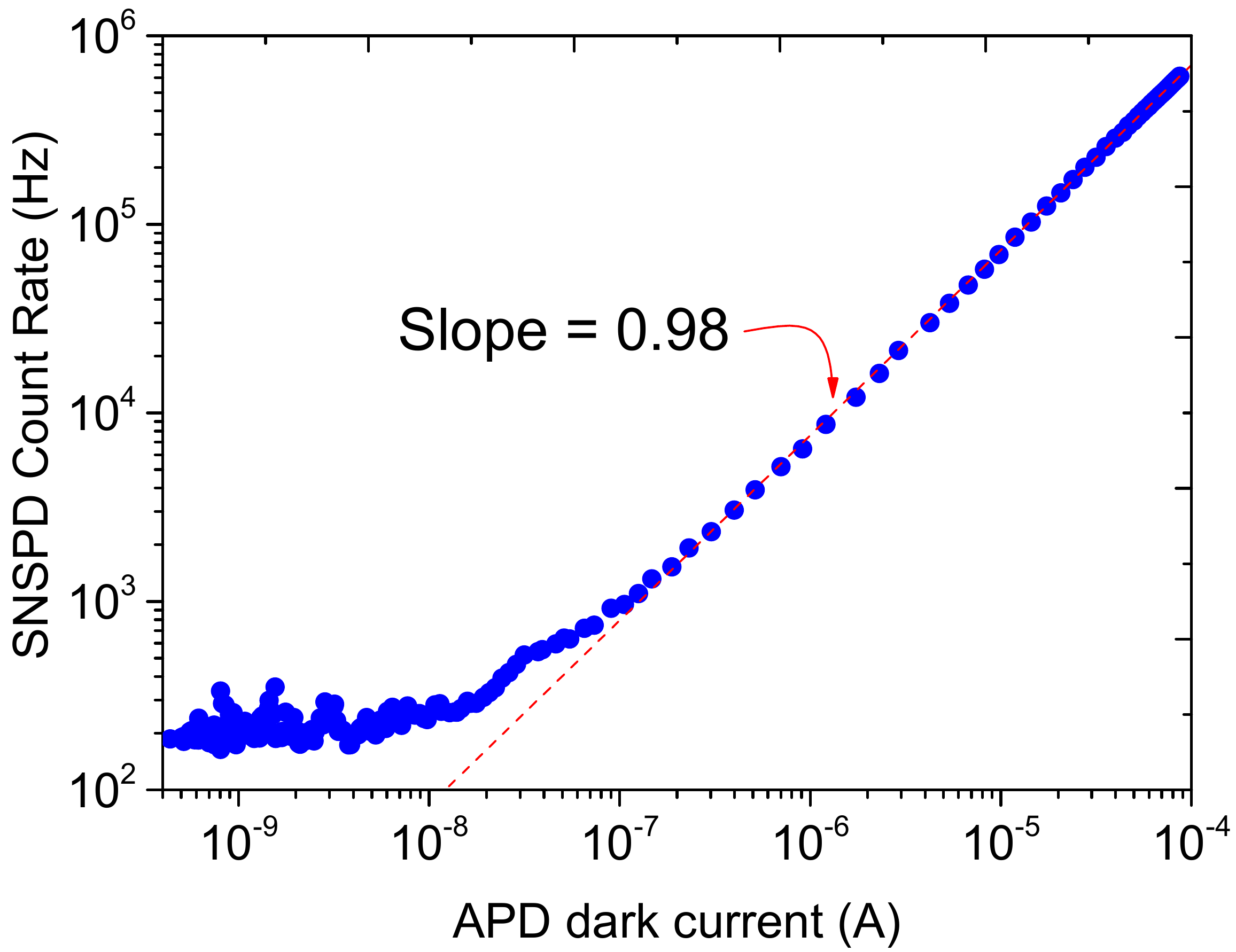}
    \caption{SNSPD count rate as a function of APD dark current. The linear relationship between the two strongly points to backflashes originating in the InP multiplication region.}
    \label{fig:backflash_dark}
\end{figure}

As a second experiment to probe the origin of the APD backflashes, we switch off the laser in and measure the backflashes with the APD kept under dark conditions. We measure the SNSPD count rate as a function of the APD dark current by adjusting the DC bias to the APD. The result is given in Fig.~\ref{fig:backflash_dark}.

Initially the SNSPD count rate remains at the dark count level until the APD current reaches a value of approximately 10 nA. After about 100 nA, the data appears to follow a linear trend and this is confirmed by fitting the data points. This finding supports the hypothesis that backflashes arise from carriers in the multiplication region; a higher dark current arises from the larger electric field increasing the avalanche probability, thereby generating more carriers which cause backflashes.

In conclusion, we have investigated backflashes in GHz-gated self-differencing InGaAs APDs. By performing the first characterisation of the backflash rate in these devices using high efficiency SNSPDs, we have found evidence that supports the hypothesis that shorter gates lead to fewer backflashes. We have shown that the information leakage as a result of backflashes has a negligible effect on the secure key rate in QKD and is, as such, of minimal concern in QKD systems. We have performed characterisation indicates backflashes originate in the detector's InP multiplication region.

\begin{acknowledgments}
A. K.-S. gratefully acknowledges financial support from Toshiba Research Europe Ltd and the Engineering and Physical Sciences Research Council through an Industrial CASE studentship.
\end{acknowledgments}



\bibliographystyle{unsrt}
\bibliography{aipsamp_bk}

\begin{thebibliography}{10}

\bibitem{bennett1984}
Charles~H Bennett and Gilles Brassard.
\newblock Quantum cryptography: Public key distribution and coin tossing.
\newblock In {\em International Conference on Computers, Systems \& Signal
  Processing, Bangalore, India, Dec 9-12, 1984}, pages 175--179, 1984.

\bibitem{Peev09}
M~Peev, C~Pacher, R~Alléaume, C~Barreiro, J~Bouda, W~Boxleitner,
  T~Debuisschert, E~Diamanti, M~Dianati, J~F Dynes, S~Fasel, S~Fossier,
  M~Fürst, J-D Gautier, O~Gay, N~Gisin, P~Grangier, A~Happe, Y~Hasani,
  M~Hentschel, H~Hübel, G~Humer, T~Länger, M~Legré, R~Lieger, J~Lodewyck,
  T~Lorünser, N~Lütkenhaus, A~Marhold, T~Matyus, O~Maurhart, L~Monat,
  S~Nauerth, J-B Page, A~Poppe, E~Querasser, G~Ribordy, S~Robyr, L~Salvail, A~W
  Sharpe, A~J Shields, D~Stucki, M~Suda, C~Tamas, T~Themel, R~T Thew, Y~Thoma,
  A~Treiber, P~Trinkler, R~Tualle-Brouri, F~Vannel, N~Walenta, H~Weier,
  H~Weinfurter, I~Wimberger, Z~L Yuan, H~Zbinden, and A~Zeilinger.
\newblock {The SECOQC quantum key distribution network in Vienna}.
\newblock {\em New Journal of Physics}, 11(7):075001, 2009.

\bibitem{Sasaki11}
M.~Sasaki, M.~Fujiwara, H.~Ishizuka, W.~Klaus, K.~Wakui, M.~Takeoka, S.~Miki,
  T.~Yamashita, Z.~Wang, A.~Tanaka, K.~Yoshino, Y.~Nambu, S.~Takahashi,
  A.~Tajima, A.~Tomita, T.~Domeki, T.~Hasegawa, Y.~Sakai, H.~Kobayashi,
  T.~Asai, K.~Shimizu, T.~Tokura, T.~Tsurumaru, M.~Matsui, T.~Honjo, K.~Tamaki,
  H.~Takesue, Y.~Tokura, J.~F. Dynes, A.~R. Dixon, A.~W. Sharpe, Z.~L. Yuan,
  A.~J. Shields, S.~Uchikoga, M.~Legr\'{e}, S.~Robyr, P.~Trinkler, L.~Monat,
  J.-B. Page, G.~Ribordy, A.~Poppe, A.~Allacher, O.~Maurhart, T.~L\"{a}nger,
  M.~Peev, and A.~Zeilinger.
\newblock Field test of quantum key distribution in the {Tokyo QKD Network}.
\newblock {\em Opt. Express}, 19(11):10387--10409, May 2011.

\bibitem{Dynes12}
J.~F. Dynes, I.~Choi, A.~W. Sharpe, A.~R. Dixon, Z.~L. Yuan, M.~Fujiwara,
  M.~Sasaki, and A.~J. Shields.
\newblock Stability of high bit rate quantum key distribution on installed
  fiber.
\newblock {\em Opt. Express}, 20(15):16339--16347, Jul 2012.

\bibitem{Mao:18}
Yingqiu Mao, Bi-Xiao Wang, Chunxu Zhao, Guangquan Wang, Ruichun Wang, Honghai
  Wang, Fei Zhou, Jimin Nie, Qing Chen, Yong Zhao, Qiang Zhang, Jun Zhang,
  Teng-Yun Chen, and Jian-Wei Pan.
\newblock Integrating quantum key distribution with classical communications in
  backbone fiber network.
\newblock {\em Opt. Express}, 26(5):6010--6020, Mar 2018.

\bibitem{Dixon17}
A.~R. Dixon, J.~F. Dynes, M.~Lucamarini, B.~Fr\"{o}hlich, A.~W. Sharpe,
  A.~Plews, W.~Tam, Z.~L. Yuan, Y.~Tanizawa, H.~Sato, S.~Kawamura, M.~Fujiwara,
  M.~Sasaki, and A.~J. Shields.
\newblock Quantum key distribution with hacking countermeasures and long term
  field trial.
\newblock {\em Sci. Rep.}, 7(1):1978, 2017.

\bibitem{Bunandar18}
Darius Bunandar, Anthony Lentine, Catherine Lee, Hong Cai, Christopher~M. Long,
  Nicholas Boynton, Nicholas Martinez, Christopher DeRose, Changchen Chen,
  Matthew Grein, Douglas Trotter, Andrew Starbuck, Andrew Pomerene, Scott
  Hamilton, Franco N.~C. Wong, Ryan Camacho, Paul Davids, Junji Urayama, and
  Dirk Englund.
\newblock Metropolitan quantum key distribution with silicon photonics.
\newblock {\em Phys. Rev. X}, 8:021009, Apr 2018.

\bibitem{Sun18}
Wei Sun, Liu-Jun Wang, Xiang-Xiang Sun, Yingqiu Mao, Hua-Lei Yin, Bi-Xiao Wang,
  Teng-Yun Chen, and Jian-Wei Pan.
\newblock Experimental integration of quantum key distribution and
  gigabit-capable passive optical network.
\newblock {\em Journal of Applied Physics}, 123(4):043105, 2018.

\bibitem{Comandar14}
L.~C. Comandar, B.~Fr\"{o}hlich, M.~Lucamarini, K.~A. Patel, A.~W. Sharpe,
  J.~F. Dynes, Z.~L. Yuan, R.~V. Penty, and A.~J. Shields.
\newblock Room temperature single-photon detectors for high bit rate quantum
  key distribution.
\newblock {\em Appl. Phys. Lett.}, 104(2):021101, 2014.

\bibitem{Comandar16}
L.~C. Comandar, M.~Lucamarini, B.~Fröhlich, J.~F. Dynes, A.~W. Sharpe,
  S.~W.-B. Tam, Z.~L. Yuan, R.~V. Penty, and A.~J. Shields.
\newblock Quantum key distribution without detector vulnerabilities using
  optically seeded lasers.
\newblock {\em Nature Photonics}, 10(5):312--315, April 2016.

\bibitem{Yuan18}
Zhiliang Yuan, Alan Plews, Ririka Takahashi, Kazuaki Doi, Winci Tam, Andrew
  Sharpe, Alexander Dixon, Evan Lavelle, James Dynes, Akira Murakami, Mamko
  Kujiraoka, Marco Lucamarini, Yoshimichi Tanizawa, Hideaki Sato, and Andrew~J.
  Shields.
\newblock {10-Mb/s Quantum Key Distribution}.
\newblock {\em J. Lightwave Technol.}, 36(16):3427--3433, Aug 2018.

\bibitem{Makarov05}
Vadim Makarov and Dag~R. Hjelme.
\newblock Faked states attack on quantum cryptosystems.
\newblock {\em J. Mod. Opt.}, 52(5):691--705, 2005.

\bibitem{Lydersen_hacking_10}
Lars Lydersen, Carlos Wiechers, Christoffer Wittmann, Dominique Elser, Johannes
  Skaar, and Vadim Makarov.
\newblock Hacking commercial quantum cryptography systems by tailored bright
  illumination.
\newblock {\em Nat. Photon.}, 4(10):686--689, oct 2010.

\bibitem{Gerhardt11}
Ilja Gerhardt, Qin Liu, Antía Lamas-Linares, Johannes Skaar, Christian
  Kurtsiefer, and Vadim Makarov.
\newblock Full-field implementation of a perfect eavesdropper on a quantum
  cryptography system.
\newblock {\em Nature Communications}, 2:349, jun 2011.

\bibitem{Yuan10}
Z.~L. Yuan, J.~F. Dynes, and A.~J. Shields.
\newblock Avoiding the blinding attack in {QKD}.
\newblock {\em Nat. Photon.}, 4(12):800--801, 2010.

\bibitem{Yuan11}
Z.~L. Yuan, J.~F. Dynes, and A.~J. Shields.
\newblock Resilience of gated avalanche photodiodes against bright illumination
  attacks in quantum cryptography.
\newblock {\em Appl. Phys. Lett.}, 98(23):231104, 2011.

\bibitem{Meda17}
Alice Meda, Ivo~Pietro Degiovanni, Alberto Tosi, Zhiliang Yuan, Giorgi Brida,
  and Marco Genovese.
\newblock Quantifying backflash radiation to prevent zero-error attacks in
  quantum key distribution.
\newblock {\em Light: Science and Applications}, 6(e16261), 2017.

\bibitem{Pinheiro18}
Paulo Vinicius~Pereira Pinheiro, Poompong Chaiwongkhot, Shihan Sajeed, Rolf~T.
  Horn, Jean-Philippe Bourgoin, Thomas Jennewein, Norbert L\"{u}tkenhaus, and
  Vadim Makarov.
\newblock Eavesdropping and countermeasures for backflash side channel in
  quantum cryptography.
\newblock {\em Opt. Express}, 26(16):21020--21032, Aug 2018.

\bibitem{Marini17}
Loris Marini, Robin Camphausen, Benjamin~J. Eggleton, and Stefano Palomba.
\newblock Deterministic filtering of breakdown flashing at telecom wavelengths.
\newblock {\em Applied Physics Letters}, 111(21):213501, 2017.

\bibitem{Shi17}
Yicheng Shi, Janet Zheng~Jie Lim, Hou~Shun Poh, Peng~Kian Tan, Peiyu~Amelia
  Tan, Alexander Ling, and Christian Kurtsiefer.
\newblock {Breakdown flash at telecom wavelengths in InGaAs avalanche
  photodiodes}.
\newblock {\em Opt. Express}, 25(24):30388--30394, Nov 2017.

\bibitem{Kurtsiefer01}
Christian Kurtsiefer, Patrick Zarda, Sonja Mayer, and Harald Weinfurter.
\newblock The breakdown flash of silicon avalanche photodiodes-back door for
  eavesdropper attacks?
\newblock {\em Journal of Modern Optics}, 48(13):2039--2047, 2001.

\bibitem{Comandar15}
L.~C. Comandar, B.~Fr\"{o}hlich, J.~F. Dynes, A.~W. Sharpe, M.~Lucamarini,
  Z.~L. Yuan, R.~V. Penty, and A.~J. Shields.
\newblock Gigahertz-gated {InGaAs}/{InP} single-photon detector with detection
  efficiency exceeding 55\% at 1550 nm.
\newblock {\em J. Appl. Phys.}, 117(8):083109, feb 2015.

\bibitem{Gottesman04}
Daniel Gottesman, Hoi-Kwong Lo, Norbert L\"{u}tkenhaus, and John Preskill.
\newblock {Security of Quantum Key Distribution with Imperfect Devices}.
\newblock {\em Quantum Info. Comput.}, 4(5):325--360, sep 2004.

\bibitem{Inamori07}
H.~Inamori, N.~L{\"u}tkenhaus, and D.~Mayers.
\newblock Unconditional security of practical quantum key distribution.
\newblock {\em The European Physical Journal D}, 41(3):599, Jan 2007.

\end{thebibliography}

\end{document}